\documentclass[aps,twocolumn,showpacs]{revtex4}
\usepackage{epsfig}
\newcommand{\be}{\begin{equation}}
\newcommand{\ee}{\end{equation}}
\newcommand{\bea}{\begin{eqnarray}}
\newcommand{\eea}{\end{eqnarray}}
\begin{document}

\title{The average shape of a fluctuation: universality in excursions
of stochastic processes}
\author{Andrea Baldassarri}
\email{Andrea.Baldassarri@roma1.infn.it}
\author{Francesca Colaiori}
\email{fran@pil.phys.uniroma1.it}
\author{Claudio Castellano}
\email{castella@pil.phys.uniroma1.it}
\affiliation{INFM UdR Roma 1 - Dipartimento di Fisica, 
Universit\`a ``La Sapienza", P.le A. Moro 2,
00185 - Roma, Italy.}

\date{\today}

\begin{abstract}
We study the average shape of a fluctuation of a time series $x(t)$,
that is the average value $\langle x(t)-x(0) \rangle_T$
before $x(t)$ first returns, at time $T$, to its initial value $x(0)$.
For large classes of stochastic processes we find that a scaling
law of the form $\langle x(t) - x(0) \rangle_T = T^\alpha f(t/T)$ is obeyed.
The scaling function $f(s)$ is to a large extent independent of the
details of the single increment distribution,
while it encodes relevant statistical information
on the presence and nature of temporal correlations in the process.
We discuss the relevance of these results for Barkhausen noise
in magnetic systems.
\end{abstract}

\maketitle

Many experimental investigations of a wide range of complex systems
involve the measure of some scalar observable over long time
intervals, during which the signal exhibits nontrivial fluctuations
around some average value or avalanche-like bursts of activity
separated by quiescent intervals~\cite{Sethna01}.  A few examples are
Barkhausen noise in magnetic materials~\cite{Bertottibook}, solar
flares in astrophysics~\cite{Lu}, seismic activity in
geophysics~\cite{Kanasewichbook}, or prices in financial
markets~\cite{Bouchaudbook}.  The statistical features of such
fluctuations reflect the properties of the dynamics that generates
them, and their analysis is a key point for understanding the system
under investigation. Traditional tools for characterizing temporal
series are correlation functions, distributions of durations of
fluctuations and of their amplitudes.  Here we focus on the average
shape of a fluctuation, showing that it contains crucial pieces of
information about the nature of the underlying process.  Since this
quantity can be easily extracted from experimental data, it is an
useful tool in the statistical analysis of temporal series.

The average shape of a fluctuation has been lately the subject of
interest in the study of the Barkhausen
effect~\cite{Sethna01,Kuntz00,Durin02}, where it has been used as a
stringent test for the validity of theories against experiments.  All
current models, successfully used to describe most of the features of
the phenomenon, are however unable to reproduce the form of
fluctuations (pulses) measured experimentally~\cite{Mehta01}.  In
particular, while all models predict a shape symmetric with respect to
its middle point, empirical data yield a skewed form.  This indicates
that the understanding of the Barkhausen noise is still
incomplete. Moreover it shows that the average shape of a fluctuation
is a sharper tool for discriminating universality classes than
critical exponents~\cite{Sethna01}.  Motivated by these recent
results, we approach theoretically the problem of finding the average
shape of a fluctuation for a generic signal, and of understanding
which statistical features of the process generating the signal are
encoded in this curve.

\begin{figure}
\centerline{\psfig{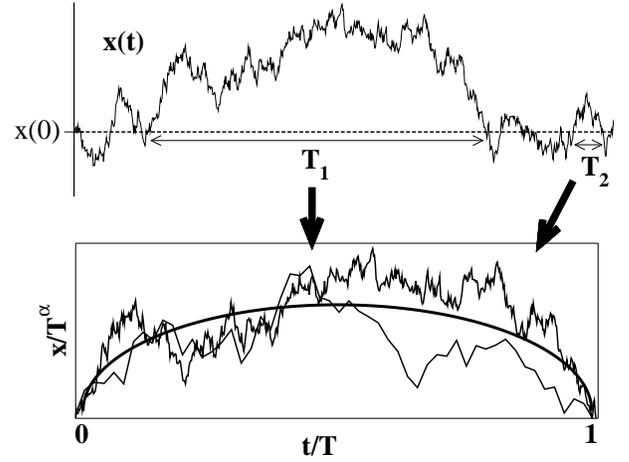}}
\caption{Schematic illustration of the definition of the average shape of a 
fluctuation. Above, the signal $x(t)$ is shown and two successive
positive fluctuations above $x(0)$ are indicated.
Below the two fluctuations are rescaled and compared to the average shape.
}
\label{Fig0}
\end{figure}

Let us define precisely what we intend by average fluctuation of a
generic signal (Fig.~\ref{Fig0}).
If $x(t)$ is the amplitude of a signal, we define as duration $T$ of the
pulse the time needed for $x(t)$ to return for the first time to $x(0)$.
Collecting positive pulses of the same duration we introduce the average
amplitude of the signal after time $t$ elapsed from the beginning of a
fluctuation
\[
\langle x(t)-x(0) \rangle_T.
\]
Interpreting $x(t)$ as the position of a one-dimensional walker at
time $t$, $\langle x(t)-x(0) \rangle_T$ is the average trajectory of
the walker before it returns to the starting position $x(0)$, and is
obtained by summing over all positive walks starting at time $t=0$ in
$x=x(0)$ and constrained to return for the first time to $x(0)$ at
time $t=T$.

In this Letter we compute this average trajectory for stochastic
processes of the form
\be
x(t+1)=x(t)+\xi(t)
\label{randomwalk}
\ee
with $x(0)=0$~\cite{nota}.
The increment $\xi(t)$ is a random variable.  Despite the simplicity of
the processes considered, we uncover a rich phenomenology.  We find
that, in all cases considered, a scaling law is obeyed
\be
\langle
x(t) \rangle_T = T^\alpha f(t/T),
\label{scaling}
\ee
in appropriate time regimes.
The value of the exponent $\alpha$ results to be equal to the wandering
exponent of the ``free'' process (i. e. not constrained to return to 0 at
time $T$, nor to be always positive).
The scaling function $f(s)$ turns out instead to be a sensitive probe of
temporal correlations of the dynamical process, while it is to a large extent
independent on the statistics of single increments.
In particular, for uncorrelated noise the scaling function
is always proportional to a semicircle. For correlated noise the tails 
are universal with respect to the single step distribution, and they 
depend only on the short- or long-ranged nature of the correlations.

We consider processes of the form (\ref{randomwalk}) and focus our 
attention on the trajectory of a walk originating in $x(0)=0$ at time
$0$ and returning to $x(0)$ for the first time at time $T$,
known in the mathematical literature as {\it excursion}. 
We first study the case of uncorrelated random walks and 
Levy flights and then move to the analysis of the effect of correlations.    

{\em i) Gaussian walks:}
The simplest process of this kind is the uncorrelated unbiased random walk,
where $\langle \xi(t) \rangle=0$ and $\langle \xi(t) \xi(t')
\rangle=\delta_{t,t'}$.
By using the time-reversal symmetry the average trajectory can be written as
\be
\langle x(t) \rangle_T = {\int_0^{\infty} dx \, \, x \, \,F(x,t) F(x,T-t) \over
\int_0^{\infty} dx \, \, F(x,t) F(x,T-t)}.
\label{langrang}
\ee
where $F(x,t)$ is the probability that a walker starting in $x=0$ at
time $0$ reaches $x$ at time $t$ without ever touching the axis $x=0$.

If we consider a normal distribution for single steps $F(x,t)$
can be determined via the image method~\cite{Rednerbook}.
The result for $\langle x(t) \rangle_T$ is of the form~(\ref{scaling})
with $\alpha=1/2$ and scaling function proportional to a
semicircle~\cite{long}
\be
f_U(s) = \sqrt{8 \over \pi} \sqrt{s(1-s)}.
\label{semicircle}
\ee
The exponent $\alpha$, being equal to the value characterizing a free
walk, indicates that the constraint of returning at time $T$ does not
affect the amplitude of the excursion. 
The same value of $\alpha$ and the same expression~(\ref{semicircle})
for $f(s)$ are expected for any distribution of single steps $P(\xi)$
with finite variance, for which central limit theorem holds.
Eq.~(\ref{semicircle}) can be easily computed explicitly
in the case of a bimodal $P(\xi)$.

To extend our analysis to other processes we resort
to numerical simulations for the evaluation of $\langle x(t) \rangle_T$.
We consider walks starting at $x=0$ at some negative time,
and we take as $t=1$ the first time such that $x(t)> \epsilon$, where
$\epsilon$ is a small positive threshold.
We average over all walks that first return between $[-\epsilon,\epsilon]$
at a specified time $T$, under the constraint $x(t)>\epsilon$ for $1<t<T$.
For each time $T$ the average shape is normalized by the factor
$N(T)=\int_{0}^1 ds \langle x(sT) \rangle_T$.
If the scaling form~(\ref{scaling}) holds, the normalized shapes for
different $T$ collapse on a single curve and $N(T)$ grows as
$T^{\alpha}$.
We average from $10^4$ to $10^7$ pulses for each $T$.
However, as shown in experiments~\cite{Mehta01},
smaller statistics ($\approx 10^3$ pulses) is already sufficient
for reasonably clean curves.

\begin{figure}
\centerline{\psfig{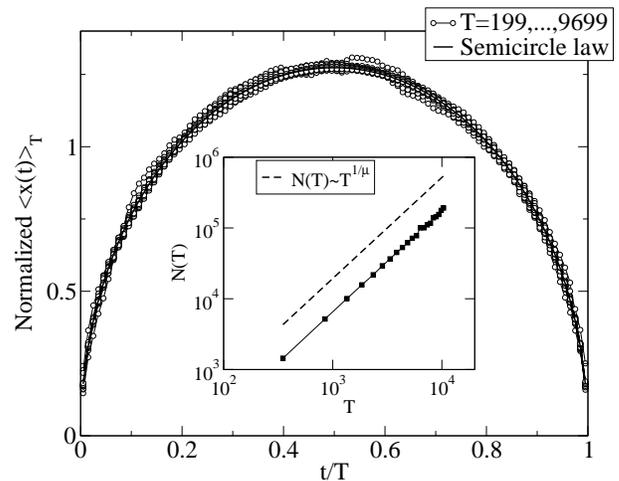}}
\caption{Main: Plot of the normalized average fluctuation
$\langle x(t)\rangle_T$ for a random walk with power-law distributed steps
with $\mu=0.7$. The solid line represents the universal
function~(\ref{semicircle}). Small circles are for several values of $T$
ranging from 199 to 9699. Inset: Plot of the normalization factor $N(T)$,
which scales as $T^{1/\mu}$, indicating that $\alpha=1/\mu$.
The dashed line is proportional to $T^{1/\mu}$.}
\label{Fig1}
\end{figure}

{\em ii) Levy flights:}
While it is well known that random walks with finite variance of the
step distribution belong to the Gaussian universality class,
distributions with fat power-law tails give rise to a completely
different behavior~\cite{Bouchaud90}.
For this reason we consider now process~(\ref{randomwalk}) where the
distribution of independent increments $\xi$ has infinite variance,
i.e. with tails decaying as $P(\xi) \propto |\xi|^{-\mu-1}$ where
$0 < \mu < 2$.
Although $\langle x(t)\rangle_T$ can still be written in the
form~(\ref{langrang}), $F(x,t)$ can not be computed by the image method.
We find numerically that also in this case
the scaling form~(\ref{scaling}) is obeyed, 
with $\alpha=1/\mu$ and the scaling function $f(s)$ of the same
semicircular shape~(\ref{semicircle}) of the case with finite variance.
This holds for any $\mu>0$ (in Fig.~\ref{Fig1} we show 
$\langle x(t)\rangle_T$ for $\mu=0.7$), thus even regardless the existence
of the first (absolute) moment of the single step distribution
\cite{notaCauchy}.

While a scaling exponent $\alpha$ equal to the wandering exponent
of the free process could be expected~\cite{notaalpha},
the fact that the shape of the scaling function remains exactly the same is
surprising.
Note that the short time expansion
of Eq.~(\ref{scaling})
\be
\langle x(t)\rangle_T \propto  T^{1/\mu-1/2} \, \, t^{1/2}
\label{short}
\ee
in the Gaussian case $\mu=2$ gives $\langle x(t)\rangle_T \propto  t^{1/2}$, 
equal to the wandering of the unconstrained walk.
One could naively expect by analogy a short time
behavior as $t^{1/\mu}$ for $\mu<2$ (corresponding to the free walk).
Eq.~(\ref{short}) shows that this is not the case, indicating that
the Levy flight feels the constraints even for short times.

Universality in the shape of the scaling function is not restricted to changes
of the variance $\langle \xi^2 \rangle$ of the single step distribution.
Also variations of the mean value $\langle \xi \rangle$, corresponding
to the addition of a bias to the walk, do not modify $f(s)$.
This is easily shown analytically in the case of 
gaussian or bimodal $P(\xi)$ and has been checked numerically for the
other cases.
Therefore we can conclude that, for {\em all} processes of the
type~(\ref{randomwalk}) with uncorrelated noise, the scaling function
$f(s)$ is proportional to a semicircle.

{\em iii) Walk in a parabolic well:}
A further generalization is to consider a walk in a harmonic potential
\be
x(t+1)=x(t)-\lambda x(t)+\xi(t)
\label{well}
\ee
where the term $\lambda x$ describes the effect of a parabolic well.
The damping $\lambda$ introduces a characteristic time, of order $1/\lambda$,
so that scaling breaks down for large times.
The probability for the walker to return at time $T$, i. e. the
distribution of return times $P(T)$, remains the same ($\propto T^{-3/2}$)
for $T \ll 1/\lambda$, while it decays exponentially for $T \gg 1/\lambda$.
Also for this process the average fluctuation can be computed analytically
via the image method and is~\cite{long}
\be
\langle x(t) \rangle_T 
= \sqrt{{4 \over \pi \lambda} {(1-e^{-2 \lambda t})
(1-e^{-2 \lambda (T-t)}) \over (1-e^{-2 \lambda T})}}
\label{damped}
\ee
\begin{figure}
\centerline{\psfig{file=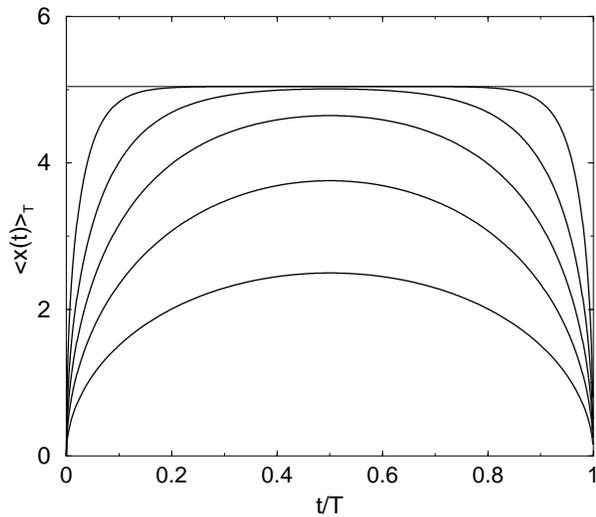,width=7cm,angle=-90,clip=!}}
\caption{
Average excursion for a random walk in a parabolic well [Eq.~(\ref{damped})]
for $1/\lambda=20$.
From top to bottom lines are for $T=10, 25, 50, 100, 250$.
Notice that the shape flattens to the constant value
$\sqrt{4/(\pi \lambda)}$ (thin line).}
\label{Fig2}
\end{figure}
As shown in Fig.~\ref{Fig2}, for $T \ll 1/\lambda$ we get a semicircle,
the same result as for $\lambda=0$.
For larger $T$ instead, as correlations decay, the form of the
fluctuation tends to become flat, while keeping the $(t/T)^{1/2}$
($(1-t/T)^{1/2}$) behavior at the left (right) tail.
This is strongly reminiscent of the flattening observed experimentally by
Durin and Zapperi in the study of the Barkhausen noise in magnetic
systems~\cite{Durin02b}:
indeed, in that case the damping term is provided the demagnetizing field,
which controls the cutoff in the avalanche distribution~\cite{Zapperi98}.

So far we have considered processes with increments $\xi(t)$ extracted
independently at each time step.
Let us now analyze the effect of temporal correlations in the
stochastic noise. More precisely, we consider processes such that
\be
g(t,t') \equiv \langle \xi(t) \xi(t') \rangle - \langle \xi(t) \rangle
\langle \xi(t') \rangle \neq \delta_{t,t'}.
\ee

\begin{figure}[t]
\centerline{\psfig{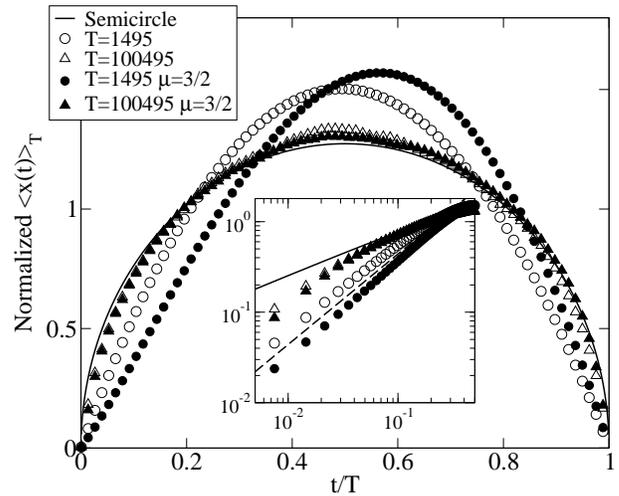}}
\caption{
Main: Plot of the normalized average fluctuation
$\langle x(t)\rangle_T$ for a random walk fed by noise
with short time memory ($\tau=1000$).
Empty symbols are for uniformly distributed
$\eta$, while the filled ones are for power-law distributed $\eta$
with $\mu=3/2$. Circles are for a time of the order of $\tau$,
while triangles are for $T \gg \tau$.
The solid line represent the universal function~(\ref{semicircle})
after normalization.
Inset: The tails for small $t/T$ of the same curves of the main panel
plus a straight dashed line of slope 1.}
\label{Fig3}
\end{figure}

{\em iv) Short-ranged memory:}
Let us first study the case of noise with correlations decaying
exponentially over some interval $\tau$, i.e.
$g(t,t')=\exp{(-|t-t'|/\tau)}$.
The average fluctuation for this process, evaluated numerically,
is reported in Fig.~\ref{Fig3}, showing the existence of two regimes.
When $T\approx \tau$ the shape of $\langle x(t)\rangle_T$ is parabolic, 
similarly to what has been observed in \cite{Mehta01} for the random field 
Ising model.
The detailed form depends on the distribution of single steps $P(\xi)$
and in particular it is skewed for power-law distributed $\xi$.
However, some form of universality with respect to $P(\xi)$
is present also in this case: The tails of
$\langle x(t) \rangle_T$, which go to zero linearly both
for small $s$ [Fig.~\ref{Fig3}, inset] and for $s \to 1$ (not shown).
In the long time limit $T\gg \tau$, memory in the noise is lost and,
as expected, we recover  the semicircular shape (Fig.~\ref{Fig3}),
that characterizes processes with uncorrelated increments.

\begin{figure}[b]
\centerline{\psfig{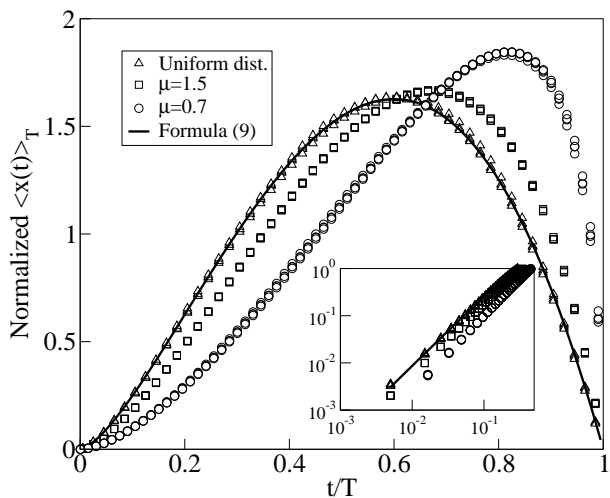}}
\caption{
Main: Plot of the normalized average fluctuation
$\langle x(t)\rangle_T$ for the random accelerated particle.
Triangles are for uniformly distributed
$\eta$, while squares and circles are for power-law distributed
$\eta$, respectively  with $\mu=1.5$ and $\mu=0.7$.
The values of $T$ range from 1000 to 100000.
Inset: The tails for small $t/T$ of the same curves of the main panel
plus a straight line of slope 3/2.}
\label{Fig4}
\end{figure}

{\em v) Long-ranged memory:}
We finally analyze a stochastic process with long-ranged memory, given
by Eq.~(\ref{randomwalk}), where now $\xi(t)=\sum_{i=1,t} \eta(i)$
and $\eta(i)$ is an uncorrelated noise.
This process can be seen as a temporally discrete version of the continuous
process $\partial_t^2 x = \eta(t)$,
which is known as the random accelerated particle (RAP) and has
been studied in the context of inelastic collapse of
granular matter~\cite{Cornell98}.
For a distribution of $\eta(t)$ with unitary variance
the correlation function of $\xi(t)$ is $g(t,t')=\min(t,t')$
and hence it does not decay when the difference between $t$ and $t'$ grows.
The first return probability to the origin of the walker
has been recently shown analytically to decay
as $P(T)\propto T ^{-5/4}$~\cite{Sinai92}.
By evaluating numerically the average fluctuation $\langle x(t) \rangle_T$
we recover also in this case the  scaling form [Eq.~(\ref{scaling})], with
$\alpha=3/2$, that again corresponds to the wandering exponent for the
unconstrained process.
The normalized scaling function, plotted in Fig.~\ref{Fig4},
is indistinguishable from the simple curve
\be
f(x)={x^{3/2}(1-x) \over B(3/2,1)}
\ee
where $B$ is the Beta function.
For a RAP with power-law distributed $\eta$
(infinite variance), the exponent $\alpha$ is $1+1/\mu$ (again as in the
free case).
As in the case with short-ranged correlations the scaling function differs
from the case with finite variance (Fig.~\ref{Fig4}), and the asymmetry
is more evident for small $\mu$.
Nevertheless the behavior of the tails remains universal, being
$f(s) \propto s^{3/2}$ for $s \to 0$ (see Fig.~(\ref{Fig4}), inset)
and $f(s)\propto 1-s$ for $s \to 1$ (not shown).

In conclusion, we have tackled theoretically the problem of analyzing
generic time series from the point of view of the average shape of
their fluctuations.  In particular, we have studied this quantity for
stochastic processes of the form~(\ref{randomwalk}).  We find that the
scaling law $\langle x(t) \rangle_T = T^\alpha f(t/T)$ is obeyed in
an appropriate time regime.
The exponent $\alpha$ coincides with the wandering exponent of the
unconstrained process, as can be expected on the basis of scaling
arguments. More interestingly, the scaling function $f(s)$ exhibits
some degree of universality.  For any process with uncorrelated
increments $f(s)$ is proportional to a semicircle, with no dependence
whatsoever on the distribution of single steps.  Modification of this
shape can be obtained only via additional $x$-dependent terms in
Eq.~(\ref{randomwalk}) or introducing correlations in the stochastic
increments.  In this last case, the average fluctuation is in general
asymmetric and, although its detailed shape depends on the distribution
of single increments,  the tails are universal, depending only on the
short- or long-ranged nature of noise correlations.
Because different universal behaviors encode solely correlation properties of
the signal, this quantity is a powerful tool in the statistical analysis of
temporal series:
In many systems it would be very interesting to compare quantitatively
$\langle x(t) \rangle_T$ between experiments and models
in the light of our results.
In the case of Barkhausen noise, where this comparison has pointed out the
limitations of current models,
our study provides a guide for refining them in order to better
capture the physics of the phenomenon.

We are grateful to Lorenzo Bertini for an interesting discussion,
to Alan Bray for useful correspondence, and to Stefano Zapperi for
a critical reading of the manuscript.
F. C. is supported by the INFM PAIS-G project ``Hysteresis in disordered
ferromagnets''.


\begin{thebibliography}{99}
\bibitem{nota}
In all processes considered here
the results are easily shown to be independent from the choice of $x(0)$,
except for process~(\ref{well}), that will be discussed elsewhere~\cite{long}.
However, one may in general expect that changing $x(0)$ will affect the
average excursion as it does for other statistical properties, e.g. the
first-return probability.
\bibitem{notaCauchy}
To corroborate the results presented in Fig. \ref{Fig1}
we have computed numerically $F(x,t)$ for the Cauchy distribution,
which is the Levy-stable process for $\mu=1$. 
The result agrees extremely well with the ansatz
$(A/t) \sqrt{x/t}/[1+(x/t)^{5/2}]$, where $A$ is a normalization constant.
Interestingly, using this form for $F(x,t)$ in Eq.~(\ref{langrang})
and performing the integrals, we recover
$\langle x(t) \rangle_T \propto T f_U(t/T)$.
\bibitem{notaalpha}
For all processes such that the average trajectory $\langle x(t) \rangle_T$
can be written in the form Eq.~(\ref{langrang}), it is enough to assume a
scaling form $t^\beta h(x/t^\alpha)$ for the conditional probability
$F(x,t)$ to have $\langle x(t) \rangle_T \propto T^{\alpha} f(t/T)$.
\end{thebibliography}
\end{document}